\documentclass[aps,prd,a4paper,twocolumn,showpacs,nofootinbib]{revtex4-1}

\usepackage{amssymb,amsmath,latexsym,mathrsfs}
\usepackage{graphicx,subfigure}
\usepackage{epsfig}
\usepackage{varioref,xr-hyper}
\usepackage{color}

\newcommand{\Mpc}{{\rm ~Mpc}}
\newcommand{\mks}{\rm ~km/s/Mpc}
\newcommand{\neff}{N_{\rm {eff}}}

\begin{document}

\title{The impact of a new median statistics $H_0$ prior on the evidence for dark radiation }

\date{May 2012\ \ KSUPT-12/1}

\author{Erminia Calabrese$^{a}$}
\author{Maria Archidiacono$^{b}$}
\author{Alessandro Melchiorri$^{b}$}
\author{Bharat Ratra$^{c}$}

\affiliation{$^a$ Sub-department of Astrophysics, Denys Wilkinson Building, University of Oxford, Keble Road, Oxford, OX1 3RH, UK}
\affiliation{$^b$ Physics Department and INFN, Universit\`a di Roma ``La Sapienza'', Ple Aldo Moro 2, 00185, Rome, Italy}
\affiliation{$^c$ Department of Physics, Kansas State University, 116 Cardwell Hall, Manhattan, KS 66506, USA}

\begin{abstract}
Recent analyses that include cosmic microwave background (CMB) anisotropy 
measurements from the Atacama Cosmology Telescope and the South Pole Telescope 
have hinted at the presence of a dark radiation component at more than 
two standard deviations. However, this result depends sensitively on the 
assumption of an HST prior on the Hubble constant, where $H_0=73.8\pm2.4 \mks$ 
at $68 \%$ c.l.. Here we repeat this kind of analysis assuming
a prior of $H_0=68 \pm2.8 \mks$ at $68 \%$ c.l., derived from
a median statistics (MS) analysis of $537$ non-CMB 
$H_0$ measurements from Huchra's compilation.
This prior is fully consistent with the value of $H_0=69.7\pm2.5 \mks$ at 
$68 \%$ c.l.\ obtained from CMB measurements under assumption of the 
standard $\Lambda$CDM model.  We show that with the MS $H_0$ prior 
the evidence for dark radiation is weakened to $\sim 1.2$ standard 
deviations. Parametrizing the dark radiation component through the 
effective number of relativistic degrees of freedom $\neff$, we find 
$\neff=3.98\pm0.37$ at $68 \%$ c.l.\ with the HST prior and 
$\neff=3.52\pm0.39$ at $68 \%$ c.l.\ with the MS prior.
We also discuss the implications for current limits on neutrino masses
and on primordial Helium abundances.
\end{abstract}
\maketitle

\section{Introduction}

Recent measurements of the cosmic microwave background (CMB) radiation 
anisotropy made by the Atacama Cosmology Telescope (ACT) \cite{act}
and by the South Pole Telescope (SPT) \cite{spt} have provided valuable
general confirmation of the theoretical predictions of the shape of
the CMB anisotropy at arcminute angular scales, in the diffusion damping 
regime.

While the inclusion of these new small-scale data do not significantly 
alter the constraints on parameters of the ``standard'' $\Lambda$CDM 
cosmological model \cite{Peebles:1984ab, Ratra:2008ab}, compared to those
obtained by using the WMAP satellite CMB anisotropy data in conjunction 
with other cosmological measurements \cite{wmap7}, they can be used to 
significantly improve the constraints on those new, ``beyond-standard-model", 
parameters that mostly affect the physics of the CMB anisotropy diffusion 
damping tail.

In particular, the recent ACT and SPT data have placed new constraints
on the number of relativistic degrees of freedom, $\neff$, that
defines the physical energy density in relativistic particles today,
$\rho_{\rm rad}$, given by
\begin{equation}
     \rho_{\rm rad}=\left[ 1+ {7 \over 8} \left({4 \over 11}\right)^{4/3} 
          \neff \right ]\rho{_\gamma} \ ,
\end{equation}
where $\rho_{\gamma}$ is the energy density of the CMB photons at 
present temperature $T_{\gamma}=2.726$~{\rm K} (see, e.g., Ref.\ \cite{fixsen}). 
In the standard scenario, assuming three active massless neutrino species 
with standard electroweak interactions, the expected value is $\neff=3.046$,  
slightly larger than 3 because of non-instantaneous neutrino decoupling 
(see, e.g., Ref. \cite{mangano3046}).

The new data from ACT and SPT, jointly analysed with earlier, large-scale 
WMAP (and other) data, rule out the case of $\neff=0$ at high statistical 
significance. That is, for the first time, CMB anisotropy and large-scale
structure observations confirm the existence of neutrinos.\footnote{
Of course, cosmological big bang nucleosynthesis theory in combination 
with the observed light nuclei abundances had pointed to this earlier.}
However, these data also seem to prefer a value of $\neff \sim 4$, hinting
at the presence of an additional relativistic component (see Refs.\ 
\cite{act, spt}), over and above the three neutrino species 
in the standard model of particle physics. In particular, some of us, 
\cite{archi2011}, found $\neff = 4.08_{-0.68}^{+0.71}$ at $95\%$ confidence 
level from such an analysis and similar results are presented in 
Refs.\ \cite{knox11, zahn}.

These results are significant, since they rather strongly suggest that 
CMB anisotropy data (alone or in conjunction with other large-scale 
cosmological data) indicate the presence of some kind of ``dark radiation''
that is not seen in any other cosmological data. They have prompted the
development of many theoretical models in which $\neff$ is larger
than $3$, \cite{Speculation}.

Several non standard models related to axions or decaying particles, gravity waves,
extra dimensions and dark energy \cite{theories} can infact predict a larger value for $\neff$.

It is therefore crucial to carefully investigate this result, to see if 
it can be strengthened or weakened by, for example, considering a 
slightly different choice of data. It is well known in the literature 
(see, e.g., Refs.\ \cite{archi2011,knox11}) that $\neff$ is degenerate 
with the value of the Hubble constant $H_0$. Assuming a prior on the 
value of the 
Hubble constant is therefore a key step in the determination of $\neff$
from the data. The prior on the Hubble constant used in most recent analyses,
labeled HST, is a Gaussian one based on the results of Ref.\ \cite{hst} 
with $H_0=73.8\pm2.4 \mks$, including systematics. 

While this 3\% determination of $H_0$ is certainly impressive, one might 
wonder if a slightly different Hubble constant prior could change the 
preference for $\neff >3$. There are several indications that a different
Hubble constant prior could be more appropriate. For instance, a number of 
measurements result in a significantly lower value of $H_0$; e.g., the 
Ref.\ \cite{Tammann:2008ab} summary value is $H_0=62.3\pm4   \mks$. In
addition, a standard analysis, under the assumption of $\neff=3.046$, 
of CMB data alone is able (in a flat universe) to constrain the Hubble 
constant. Recent such analyses yield $H_0 \sim 70 \mks$, more than one 
standard deviation away from the HST value. For example, the analysis 
of ACT and WMAP7 data in Ref.\ \cite{act} gives $H_0=69.7\pm2.5 \mks$.
Clearly, there is also observational evidence for a significantly smaller
value of $H_0$ than the HST estimate. Furthermore, it is possible that 
using a prior with a lower value of $H_0$ could result in a $\neff$
determined from CMB anisotropy and other large-scale data that is 
consistent with the other cosmological $\neff$ determinations. 

There are very many measurements of $H_0$, over 550.\footnote{
See cfa-www.harvard.edu/$\sim$huchra/.}
Most recent estimates lie in the interval 60--75$\mks$, with 
error bars on some individual estimates probably being too small, 
since these measurements are mutually inconsistent (this is 
likely a consequence of underestimated systematic errors in 
some cases). Clearly, what is needed is a convincing summary 
observational estimate of $H_0$.\footnote{
And not just for the case at hand, but for many different cosmological
parameter analyses, see, e.g., Refs.\ \cite{Chen:2004ab}.}
To date, the best technique for deriving such a summary estimate 
--- that does not make use of the error bars of the individual 
measurements --- is the median statistics technique; Ref.\ 
\cite{Gott:2001ab} includes a detailed description of this 
technique.

The median statistics technique has been used to analyse a number
of cosmological data sets. These include Type Ia supernova apparent
magnitude data, to show that the current cosmological expansion
is accelerating, \cite{Gott:2001ab, Kowalski:2008ab}; CMB temperature
anisotropy data, in one of the first analyses to show that these data
were consistent with flat spatial hypersurfaces, \cite{Podariu:2001ab};
and, collections of measurements of the cosmological clustered mass
density, in one of the earliest analyses to show that this makes up
around 25--30\% of the current epoch cosmological energy budget, 
\cite{Chen:2003ab}. These successes support the idea that a median 
statistics estimate of the Hubble constant provides an accurate summary
estimate. 

The median statistics technique has been used thrice to analyse Huchra's 
list (at three different epochs). From an analysis of 331 measurements
(up to the middle of 1999), Ref.\ \cite{Gott:2001ab} found an 
median statistics summary $H_0 = 67\mks$; from 461 measurements 
(up to the middle of 2003), and from 553 measurements (up to early 2011),
Refs.\ \cite{Chen:2003cd, Chen:2011ab} both found a median 
statistics summary $H_0 = 68\mks$. While the estimated statistical
error bar (given by the scatter in the central $H_0$ values)
has decreased as the sample size has increased, the larger (and dominant)
systematic error bar (estimated from the scatter in the summary values
of $H_0$ determined by different techniques) has changed much less.

For our analyses here we estimate $H_0$ using the method of 
Ref.\ \cite{Chen:2011ab} but now excluding from the Huchra list 
of 553 measurements the 16 $H_0$ measurements derived from CMB 
data assuming $\neff=3.046$. We exclude these $16$ CMB 
measurements as we want an external and independent prior on $H_0$ 
to use in our analysis of the latest CMB datasets. From a median 
statistics analysis of the 537 non-CMB measurements we find 
$H_0 = 68 \pm 2.8 \mks$ (one standard deviation error), identical 
to that found in Ref.\ \cite{Chen:2011ab} from an analysis of the 553
measurements. In what follows we refer to the Gaussian prior based 
on this value as the median statistics (MS) $H_0$ prior.

Our goal here is to discuss the implications of assuming the
MS prior for $H_0$, instead of the usual HST prior, for
current CMB and large-scale structure parameter inference. We 
focus much of our attention on the value of $\neff$ and the 
evidence for dark radiation, but we also consider how the MS 
prior changes the estimated value of other parameters, including  
the dark energy equation of state parameter $w$ 
and the spectral index of primordial fluctuations $n_s$.

Our paper is organized as follows. In the next Section we briefly 
summarize the data analysis method we use. In Sec.\ III we present 
our results. We conclude in Sec.\ IV. 


\section{Analysis Method}

Our analysis is based on a modified version of the public
COSMOMC \cite{Lewis:2002ah} Monte Carlo Markov Chain code. We include
the following CMB data: WMAP7 \cite{wmap7}, ACBAR \cite{acbar}, 
ACT \cite{act}, and SPT \cite{spt}, including measurements up to 
maximum multipole number $l_{\rm max}=3000$. As in Ref.\ \cite{archi2011} 
we include galaxy clustering data from the SDSS-DR7 luminous red galaxy 
sample \cite{red}. Also, as discussed in the Introduction, we choose
two different priors on the Hubble constant: the median statistics (MS) 
prior of $H_0 = 68 \pm 2.8 \mks$ as well as, for comparison, the HST 
prior \cite{hst} used in previous analyses.

In the basic analysis we sample the usual seven-dimensional set of cosmological parameters,
adopting flat priors on them: the baryon and cold dark matter densities
$\Omega_{\rm b} h^2$ and $\Omega_{\rm c} h^2$, the ratio of the sound horizon 
to the angular diameter distance at decoupling $\theta$, the optical 
depth to reionization $\tau$, the scalar spectral index $n_s$, the 
overall normalization of the spectrum $A_s$ at $k=0.002\Mpc^{-1}$, 
the effective number of relativistic degrees of freedom $\neff$.

Our analysis is very similar to the one presented in Ref.\ \cite{archi2011}, 
with three changes: (i) we consider two different $H_0$ priors; (ii) we consider
an extended case where we assume massive neutrinos, we enlarge 
our parameter space varying the total mass of neutrinos $\sum{m_\nu}$;
(iii) we allow the Helium abundance $Y_p$ to vary consistently with standard BBN 
following Ref.\ \cite{spt}. This means that each theoretical 
CMB angular spectrum is computed assuming a value for $Y_p$ derived by BBN 
nucleosynthesis from the input values of $\Omega_b h^2$ and $\neff$ of the
theoretical model considered. The small uncertainty on $Y_p$ derived
from the experimental errors on the neutron half-life produces negligible
changes in the CMB angular spectra so we ignore it.
In a latter case we also vary $Y_p$ as a free parameter. 

In addition, where indicated, we also present constraints on the dark 
energy equation of state parameter $w$ (the ratio of the pressure
to energy density of the dark energy fluid), assumed to be redshift
independent, although the corresponding dark energy density is
time dependent.\footnote{
This is the widely-used XCDM parametrization of dark energy. It is 
not a complete parametrization, as it cannot describe the evolution 
of spatial inhomogeneities, nor is it an accurate approximation of 
more physically motivated time-varying dark energy models, 
\cite{Ratra:1991ab}. It is preferable to use a consistent and 
physically motivated dark energy model, e.g., that proposed in
Refs.\ \cite{Peebles:1988ab}, for such an analysis, but this 
is a much more involved undertaking, so instead we patch up the 
XCDM parametrization by assuming that the acoustic
spatial inhomogeneities travel at the speed of light. This
extended XCDM parametrization should provide reasonable (qualitative)
indications of what might be expected in a consistent, physically-motivated
model of time-varying dark energy.}  
We consider massless neutrinos, adiabatic initial conditions, and a 
spatially-flat universe.

Following Ref.\ \cite{archi2011} we account for foreground contributions 
by marginalizing over three additional amplitudes: the Sunyaev-Zeldovich 
effect amplitude $A_{SZ}$, the amplitude of clustered point sources $A_C$,
and the amplitude of Poisson-distributed point sources $A_P$. 

\section{Results}

\subsection{Neutrinos}

\begin{table*}[htb!]
\begin{center}
\begin{tabular}{|l||c||c|c||c|c|}
\hline
\hline
 Parameters & No Prior & \multicolumn{2}{|c||}{HST Prior} & \multicolumn{2}{|c|}{MS Prior}\\
\hline
& & \multicolumn{2}{|c||}{$73.8\pm2.4 \mks$} & \multicolumn{2}{|c|}{$68 \pm 2.8 \mks$} \\
\hline
\hline
$\Omega_b h^2$ & $0.02258\pm0.00050$ & $0.02248 \pm 0.00039$ & $0.02210 \pm 0.00037$ & $0.02211 \pm 0.00040$ & $0.02188 \pm 0.00036$\\
$\Omega_c h^2$ & $0.134\pm0.010$ & $0.1317 \pm 0.0080$ & $0.1142 \pm 0.0029$ & $0.1256\pm 0.0080$ & $0.1181 \pm 0.0032$\\
$\theta$ & $1.0395\pm0.0016$ & $1.0397 \pm  0.0016$ & $1.0415 \pm 0.0014$ & $1.0400\pm 0.0017$ & $1.0409 \pm 0.0014$\\
$\tau$ & $0.085\pm0.014$ & $0.084 \pm 0.013$ & $0.083 \pm 0.013$ & $0.080 \pm 0.013$ & $0.081 \pm 0.014$\\
$n_s$ & $0.984\pm0.017$ & $0.979 \pm 0.012$ & $0.9659 \pm 0.0091$ & $0.964 \pm 0.012$ & $0.9536 \pm 0.0090$\\
$\neff$ & $4.14\pm0.57$ & $3.98 \pm 0.37$ &  $3.046$ & $3.52\pm0.39$ & $3.046$\\
$\sum m_{\nu} [{\rm eV}]$ & $0.0$ & $0.0$ & $< 0.36$ & $0.0$ & $<0.60$ \\
$H_0 [\!\!\mks]$ & $75.2\pm3.6$ & $74.2 \pm 2.0$ & $69.3 \pm 1.4$ & $70.9\pm2.1$ & $66.8 \pm 1.8$\\
$log(10^{10} A_s)$ & $3.183\pm0.043$ & $3.191 \pm 0.035$ & $3.205 \pm 0.034$ & $3.219 \pm 0.036$ & $3.226 \pm 0.034$\\
$\Omega_{\rm m}$ & $0.277 \pm 0.019$ & $0.280 \pm 0.016$ & $0.284 \pm 0.017$ & $0.294 \pm 0.017$ & $0.315 \pm 0.024$\\
$\sigma_8$ & $0.882 \pm 0.033$ & $0.876 \pm 0.028$ & $0.782 \pm 0.032$ & $0.857 \pm 0.028$ & $0.757 \pm 0.043$\\
$A_{SZ}$ & $<1.4$ & $< 1.3$ & $<0.97$ & $< 1.1$ & $<0.96$\\
$A_C [{\rm \mu K^2}]$ & $<14.5$ & $<14.7$ & $<12.8$ & $< 14.1$ & $<13.1$\\
$A_P [{\rm \mu K^2}]$ & $<24.9$ & $<25.5$ & $<26.6$ & $< 26.1$ & $<26.6$\\
\hline
$\chi^2_{\rm min}$ & $7593.4$ & $7593.2$ & $7592.0$ & $7594.8$ & $7595.1$\\
\hline
\hline
\end{tabular}
\caption{Cosmological parameter values and $68 \%$ confidence level errors
assuming $\neff$ relativistic neutrinos or $\neff=3.046$ massive neutrinos. 
$95 \%$ c.l.\ upper bounds are listed for the sum of neutrino masses and 
foregrounds parameters. We also list the derived Hubble constant, the 
non-relativistic matter density parameter $\Omega_{\rm m} 
= \Omega_c + \Omega_b$, and $\sigma_8$, the amplitude of density 
inhomogeneities averaged over spheres of radius 8$h^{-1}$ Mpc, where 
$h$ is the Hubble constant in units of 100$\mks$.}
\label{standard}
\end{center}
\end{table*}

As in Ref.\ \cite{archi2011}, we compute the likelihood function in the 
seven-dimensional (or eight when massive neutrinos are considered) 
cosmological parameter space described above, and
multiply it by the prior probability distribution functions to 
derive the seven-dimensional posterior probability density distribution
function. Marginalizing this over all but one of the cosmological
parameters gives the one-dimensional posterior probability distribution
function for the parameter of interest. This one-dimensional distribution 
function is used to determine the most likely value of the parameter, 
as well as limits on it. These are listed in Table I, for three different
Hubble constant priors: a flat one (no prior); the Gaussian HST one, 
\cite{hst}; and the Gaussian MS one.
Marginalizing over only five of the cosmological parameters, we derive the
two-dimensional posterior probability density distribution function
$P(H_0, \neff)$. This is used to derive the constraint contours 
in the two-dimensional $\neff$--$H_0$ parameter space shown in Fig.\ 1, 
for the two Gaussian $H_0$ priors. 

\begin{figure}[h!]
\includegraphics[width=\columnwidth]{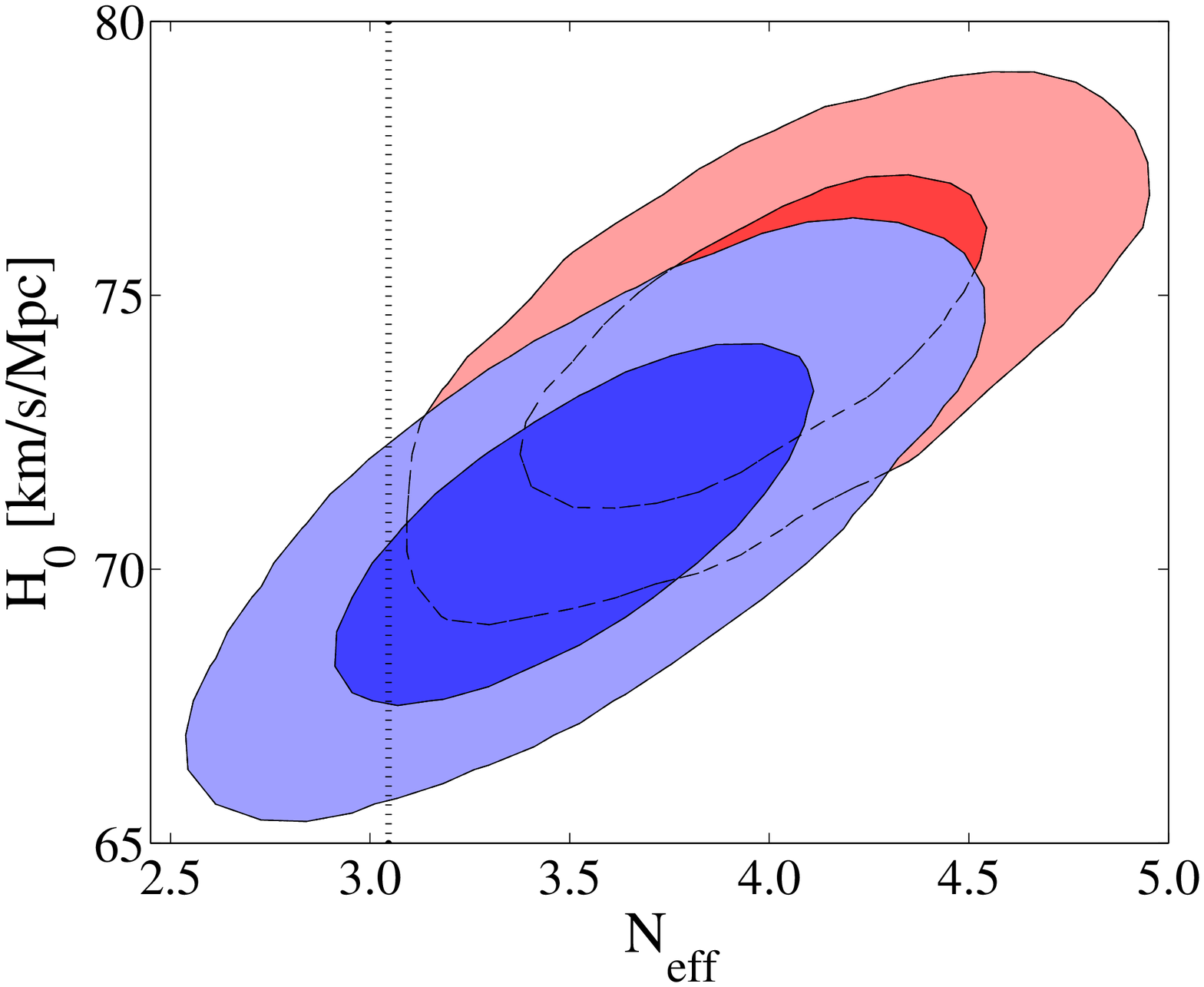}
\caption{Constraints in the $\neff$-$H_0$ plane. Elliptical two-dimensional 
posterior probability distribution function contours show the $68\%$ and 
$95\%$ c.l.\ limits. Red contours and regions (closer to the upper right 
corner) assume the HST prior with $H_0 = 73.4 \pm 2.4 \mks$, while blue 
contours and regions (closer to the lower left corner) are obtained using 
the median statistics prior with  $H_0 = 68 \pm 2.8 \mks$. The dotted black 
vertical line corresponds to $\neff = 3.046$.}  
\label{neffh0}
\end{figure}

Table I and Fig.\ \ref{neffh0} show that the $H_0$ prior plays a crucial 
role in determining constraints on $\neff$ from the data. With the HST 
$H_0$ prior we find a central $\neff$ value that is 2.5$\sigma$ larger
than 3.046, while the median statistics prior results in an $\neff$ that
is consistent with 3.046 (being only 1.2$\sigma$ larger).

The HST prior is therefore at least partially responsible for the current 
indication for dark radiation. However, as we can see from the central 
values of $H_0$ and $\neff$ obtained when a flat prior on $H_0$ is 
assumed, the CMB anisotropy and large-scale structure data considered 
here prefers a larger value of $\neff$ (being 1.9$\sigma$ larger than 
3.046) and a somewhat larger value of $H_0$. This is clear also from the 
$\chi^2_{\rm min}$ values of the best fit that are higher when the median 
statistics $H_0$ prior is assumed, compared to the case of the HST prior 
(see the last line of Table I).

\begin{figure}[h!]
\includegraphics[width=\columnwidth]{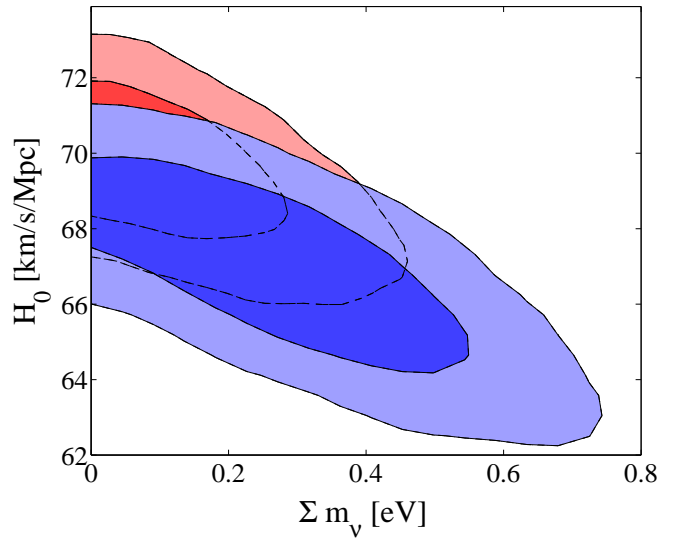}
\caption{Constraints in the $\sum{m_\nu}$-$H_0$ plane. Elliptical two-dimensional 
posterior probability distribution function contours show the $68\%$ and 
$95\%$ c.l.\ limits. Red contours and regions (closer to the upper left
corner) assume the HST prior with $H_0 = 73.4 \pm 2.4 \mks$, while blue 
contours and regions (closer to the lower right corner) are obtained using 
the median statistics prior with  $H_0 = 68 \pm 2.8 \mks$. }  
\label{mnuh0}
\end{figure}

The $H_0$ prior is crucial also in the determination of the $\sum{m_\nu}$
limits if we instead limit ourselves to the case of $3$, standard, massive 
neutrinos. 
In Table \ \ref{standard}, columns 3 and 5,
we quote the cosmological parameters and the upper limits on $\sum{m_\nu}$
in case of the HST and of the MS prior. As we can see,
the upper limit on $\sum{m_\nu}$ is considerably weaker when the MS
prior is considered, with the $95 \%$ c.l. upper limit moving from
$\sum{m_\nu} <0.36$ eV in the case of the HST prior to
$\sum{m_\nu} < 0.60$ eV in the case of the MS prior.
This can be clearly explained by the CMB degeneracy between
$H_0$ and $\sum{m_\nu}$ as illustrate in Fig.\ \ref{mnuh0}.
Namely, lower values of the Hubble parameter are in better agreement
with current CMB data when $\sum{m_\nu}$ is increased.
Dataset preferring higher values for $H_0$ will therefore provide
stronger constraints on $\sum{m_\nu}$ when combined with the 
CMB data.

Beside the $\neff$--$H_0$ degeneracy, it is interesting to note that 
there also is a degeneracy between $\neff$ and $n_s$. When the HST prior 
is assumed, $n_s$ is 1.8$\sigma$ below 1, while for the median 
statistics case it is 3$\sigma$ below unity.

\begin{figure}[h!]
\includegraphics[width=\columnwidth]{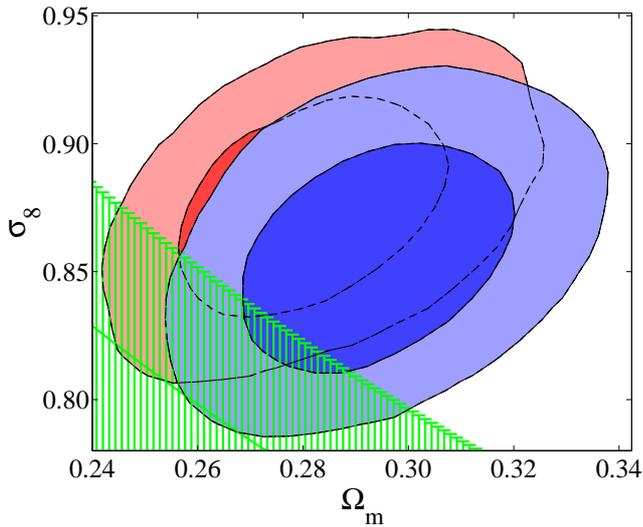}
\caption{Constraints in the $\Omega_{\rm m}$--$\sigma_8$ plane. 
Elliptical two-dimensional posterior probability density function  
contours show the $68\%$ and $95\%$ confidence level limits. Red
contours (closer to the upper left corner) assume the HST prior 
with $H_0 =73.8 \pm 2.4 \mks$; blue contours (closer to the lower
right corner) are obtained with the median statistics prior where 
$H_0 = 68 \pm 2.8 \mks$. The green region (in the lower left corner)
demarcates the central value and 2$\sigma$ limits from the cluster 
mass function normalization data, \cite{Vikhlinin:2009xb}.}
\label{sigma8_omm}
\end{figure}

In Fig.\ \ref{sigma8_omm} we show the contours in the two-dimensional 
$\Omega_{\rm m}$--$\sigma_8$ parameter space, for the two Gaussian 
$H_0$ priors. Here $\sigma_8$ is the amplitude of density 
inhomogeneities averaged over spheres of radius 8$h^{-1}$ Mpc. In 
this figure we also show the fit to the central value and the two 
standard deviation limits of the constraint from the normalization 
of the galaxy cluster mass function from Ref.\ \cite{Vikhlinin:2009xb}, 
i.e., $\sigma_8 = (0.25/\Omega_{\rm m})^{0.47} [0.813 \pm 0.013 \pm 0.024] $.
Here the first error bar represents the statistical, and the second
the systematic, error (see their Sec.\ 10). We derive the 2$\sigma$
cluster constraints shown in Fig.\ 2 by adding these errors in 
quadrature and then doubling.

From Fig.\ \ref{sigma8_omm} we see that both $H_0$ priors give results that are not
far off from what the measured normalization of the cluster mass function
demands. Qualitatively, the HST $H_0$ prior is more consistent
with the cluster data if $\Omega_{\rm m} \sim 0.25$, near the low 
end of current indications, see, e.g., Ref.\ \cite{Chen:2003ab}, while the 
median statistics case prefers a larger $\Omega_{\rm m} \sim 0.27$,
more consistent with current measurements, see, e.g., Ref.\ 
\cite{Chen:2003ab}.

\subsection{Helium mass abundance}

One assumption made in the previous paragraph is that the Helium abundance is varied
consistently with BBN. 
Current CMB data produce only weak constraints on this quantity
and allowing $Y_p$ to vary freely would make the standard case
of $\neff=3.046$ in better agreement with data due to an
anti-correlation between $\neff$ and $Y_p$ in CMB data 
(see, for example, the discussion in \cite{joudaki}).
In order to check the impact of the $H_0$ priors in this case, 
we have performed two analysis varying the Helium abundance $Y_p$ and $\neff$. 
The results are reported in Table \ref{yp}.

\begin{table}[ht!]
\begin{center}
\begin{tabular}{|l||c|c|}
\hline
\hline
Parameters & HST Prior & MS Prior\\
\hline
$\Omega_b h^2$  & $0.02274 \pm 0.00042$ & $0.02246 \pm 0.00043$ \\
$\Omega_c h^2$  & $0.1246 \pm 0.0091$ & $0.1138 \pm 0.0085$\\
$\theta$  & $1.0429 \pm 0.0027$ & $1.0454 \pm 0.0029$ \\
$\tau$  & $0.087 \pm 0.014$ & $0.085 \pm 0.014$ \\
$n_s$  & $ 0.986\pm 0.013$ & $0.972 \pm 0.013$ \\
$\neff$ & $3.52 \pm 0.48$ & $2.75 \pm 0.46$ \\ 
$H_0 [\!\!\mks]$  & $72.7 \pm 2.2$ & $68.2 \pm 2.3$ \\
$Y_p$  & $0.310 \pm 0.034$ & $0.334 \pm 0.033$\\
$log(10^{10} A_s)$  & 3.175$\pm 0.037$ & $3.197 \pm 0.036$ \\
$\Omega_{\rm m}$  & $0.279 \pm 0.015$ & $0.293 \pm 0.016$\\ 
$\sigma_8$  & $0.872 \pm 0.029$ & $0.847 \pm 0.029$ \\
$A_{SZ}$  & $<1.7$ & $<1.6$\\
$A_C [{\rm \mu K^2}]$  & $<15.4$ & $<15.3$\\
$A_P [{\rm \mu K^2}]$  & $<23.1$ & $<23.4$\\
\hline 
$\chi^2_{\rm min}$  & $7592.0$ & $7590.4$\\
\hline
\hline
\end{tabular}
\caption{Cosmological parameter values derived assuming a varying $Y_p$. 
Errors are at $68\%$ c.l.\ while upper bounds at $95 \%$ c.l. are reported 
for foregrounds parameters.}
\label{yp}
\end{center}
\end{table}

As we can see, when $Y_p$ is allowed to vary, the standard case of $\neff$ is more
consistent with current data in both cases. In the case of
the MS prior we have $\neff =2.75\pm0.46$ that is perfectly consistent with
the expectations of the standard scenario. However the value obtained for the
Helium abundance is probably too high in the case of the MS prior:
$Y_p=0.334\pm0.033$ that is about two standard deviations away from the 
conservative experimental bound of $Y_p < 0.2631$ obtained from an
analysis of direct measurements in \cite{mangaserpi}.

\begin{figure}[h!]
\includegraphics[width=\columnwidth]{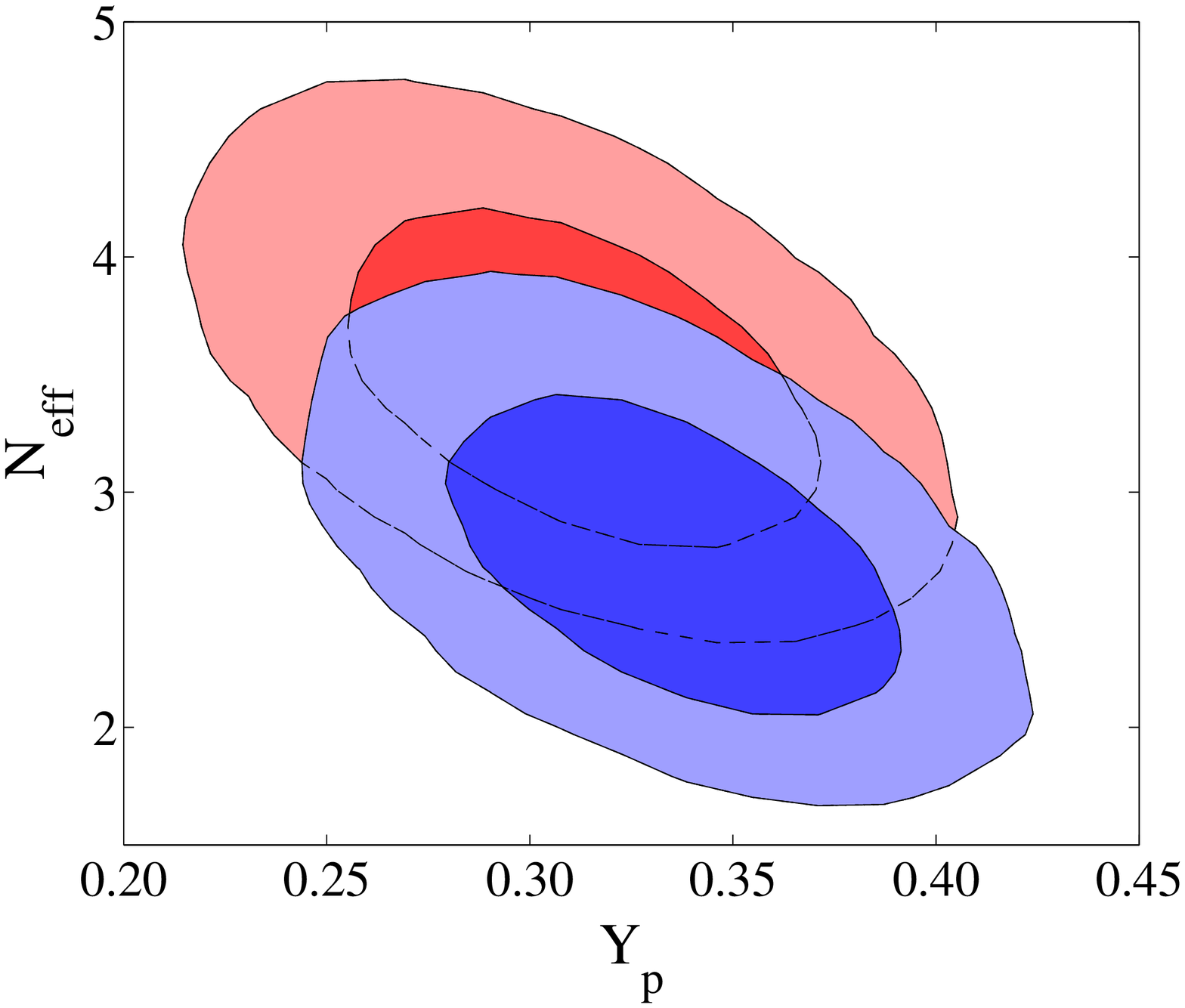}
\includegraphics[width=\columnwidth]{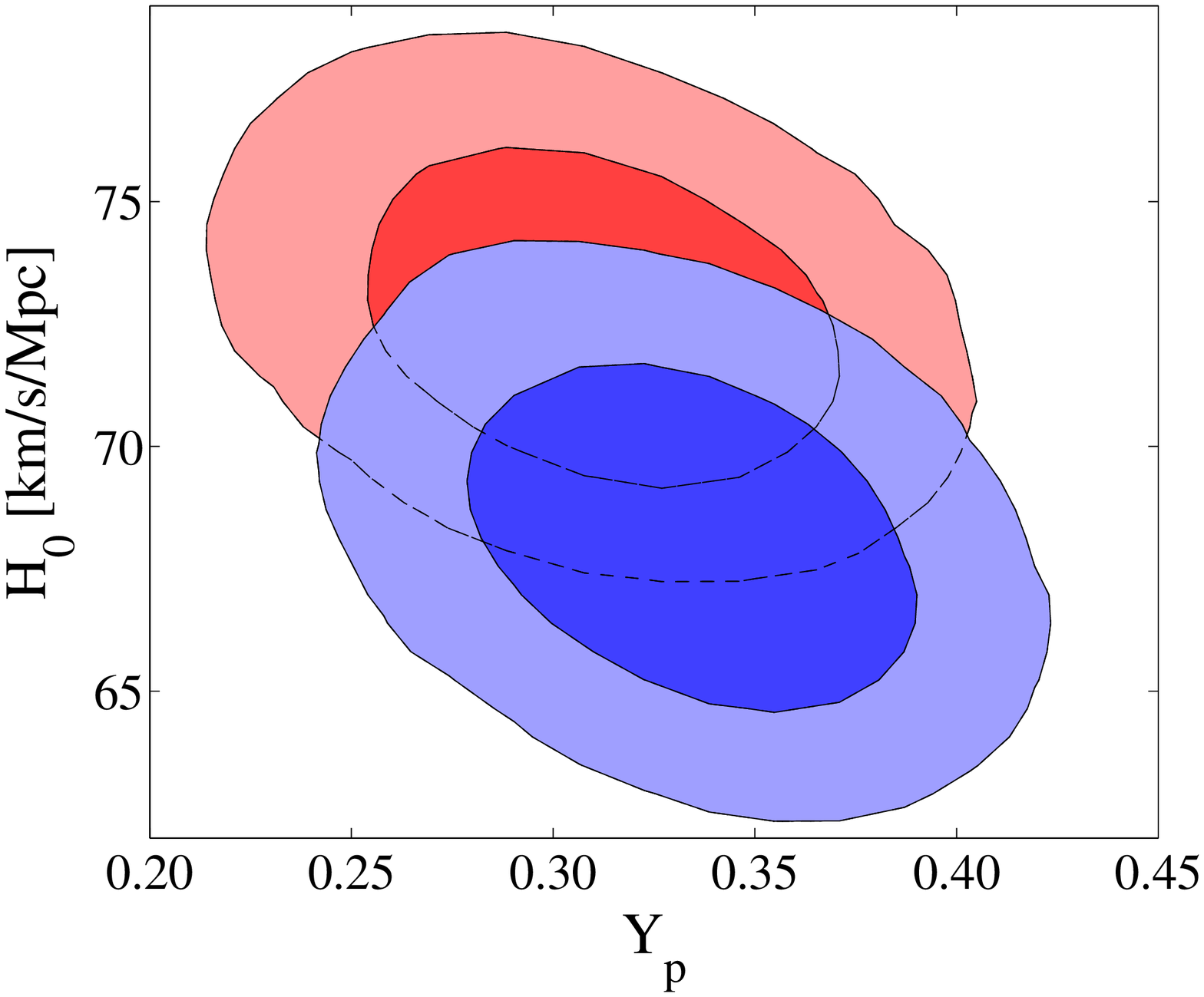}
\caption{Constraints in the $Y_p$ - $\neff$ plane (top) and  $Y_p$ -$H_0$ 
(bottom). Elliptical two-dimensional 
posterior probability distribution function contours show the $68\%$ and 
$95\%$ c.l.\ limits. Red contours and regions (closer to the upper left
corner) assume the HST prior with $H_0 = 73.4 \pm 2.4 \mks$, while blue 
contours and regions (closer to the lower right corner) are obtained using 
the median statistics prior with  $H_0 = 68 \pm 2.8 \mks$.}  
\label{ypneffh0}
\end{figure}

The larger helium abundance obtained in the case of the MS prior respect
to the HST prior can be clearly seen from the direction of the degeneracies in the 
2D contours plots in Figure \ref{ypneffh0}. Namely, 
a lower $\neff$ prefers an higher $Y_p$ and a lower prior for $H_0$ shifts 
the constraints towards lower $\neff$ and higher values for $Y_p$.

\subsection{XLCDM}

\begin{table*}[htb!]
\begin{center}
\begin{tabular}{|l||c|c||c|c|}
\hline
\hline
Parameters &\multicolumn{2}{|c||}{HST Prior}& \multicolumn{2}{|c|}{MS Prior}\\
\hline
$\Omega_b h^2$  & $0.02200\pm0.00040$ & $0.02290\pm0.0054$ & $0.02206\pm0.00040$ & $0.02279\pm0.00053$\\
$\Omega_c h^2$ & $0.1162\pm0.0039$ & $0.1347\pm0.0085$ & $0.1141\pm0.0040$ & $0.1291\pm 0.0084$\\
$\theta$ & $1.0414\pm0.0015$ & $1.0396\pm0.0016$ & $1.0414\pm0.0015$ & $1.0400\pm0.0016$\\
$\tau$ & $0.080\pm0.013$ & $0.089\pm0.015$ & $0.081\pm0.013$ & $0.089\pm0.015$\\
$n_s$ & $0.956\pm0.010$ & $0.997\pm0.019$ & $0.959\pm0.011$ & $0.993\pm0.019$\\
$\neff$ & $3.046$ & $4.42\pm0.54$ & $3.046$ & $4.16\pm0.53$\\
$H_0 [\!\!\mks]$ & $72.1\pm2.4$ & $72.8\pm2.3$ & $66.7\pm2.6$ & $68.0\pm2.4$\\
$w$ & $-1.09\pm0.10$ & $-0.86\pm0.11$ & $-0.90\pm0.10$ & $-0.76\pm 0.10$\\
$log(10^{10} A_s)$ & $3.223\pm0.039$ & $3.150\pm0.050$ & $3.210\pm0.041$ & $3.149\pm0.051$\\
$\Omega_{\rm m}$ & $0.267 \pm 0.018$ & $0.298 \pm 0.022$ & $0.307 \pm 0.226$ & $0.329 \pm 0.025$\\
$\sigma_8$ & $0.856 \pm 0.044$ & $0.831 \pm 0.047$ & $0.790 \pm 0.046$ & $0.775 \pm 0.047$\\
$A_{SZ}$ & $<0.94$ & $<1.5$ & $<0.95$ & $<1.4$\\
$A_C [{\rm \mu K^2}]$& $<13.0$ & $<15.0$ & $<13.0$ & $<14.9$\\
$A_P [{\rm \mu K^2}]$ & $<27.0$ & $<23.9$ & $<26.7$ & $<24.7$\\
\hline
$\chi^2_{\rm min}$ & $7598.1$ & $7592.7$ & $7595.1$ & $7592.1$\\
\hline
\hline
\end{tabular}
\caption{Cosmological parameter values derived assuming the XCDM 
parametrization of time-evolving dark energy. Errors are at $68\%$ c.l.\ 
while upper bounds at $95 \%$ c.l. are reported for foregrounds parameters.}
\label{wvar}
\end{center}
\end{table*}

\begin{table*}[htb!]
\begin{center}
\begin{tabular}{|l||c|c||c|c|}
\hline
\hline
Parameters &\multicolumn{2}{|c||}{HST Prior + SNeIa}& \multicolumn{2}{|c|}{MS Prior + SNeIa}\\
\hline
$\Omega_b h^2$  & $0.02203 \pm 0.00038$ & $0.02260 \pm 0.00046$ & $0.02190 \pm 0.00038$ & $0.02230 \pm 0.00046$\\
$\Omega_c h^2$ & $0.1156 \pm 0.0037$ & $0.1317 \pm 0.0079$ & $0.1157 \pm 0.0038$ & $0.1249 \pm 0.0077$\\
$\theta$ & $1.0414 \pm 0.0015$ & $1.0400 \pm 0.0016$ & $1.0411 \pm 0.0015$ & $1.0401 \pm 0.0016$\\
$\tau$ & $0.081 \pm 0.013$ & $0.086 \pm 0.014$ & $0.080 \pm 0.013$ & $0.083 \pm 0.014$\\
$n_s$ & $0.957 \pm 0.010$ & $0.985 \pm 0.015$ & $0.956 \pm 0.010$ & $0.972 \pm 0.016$ \\
$\neff$ & $3.046$ & $4.08 \pm 0.43$ & $3.046$ & $3.63 \pm 0.42$\\
$H_0 [\!\!\mks]$ & $71.0 \pm 1.6$ & $74.0 \pm 2.0$ & $68.8 \pm 1.6$ & $70.6 \pm 2.1$\\
$w$ & $-1.050 \pm 0.069$ & $-0.967 \pm 0.075$ & $-0.989 \pm 0.070$ & $-0.946 \pm 0.076$\\
$log(10^{10} A_s)$ & $3.222 \pm 0.038$ & $3.178 \pm 0.043$ & $3.221 \pm 0.038$ & $3.198 \pm 0.044$\\
$\Omega_{\rm m}$ & $0.273 \pm 0.014$ & $0.282 \pm 0.015$ & $0.291 \pm 0.015$ & $0.295 \pm 0.016$\\
$\sigma_8$ & $0.843 \pm 0.035$ & $0.863 \pm 0.038$ & $0.823 \pm 0.036$ & $0.836 \pm 0.038$\\
$A_{SZ}$ & $< 0.94$ & $< 1.3$ & $<0.94$ & $<1.2$\\
$A_C [{\rm \mu K^2}]$& $13.0$ & $14.8$ & $<13.1$ & $<14.0$\\
$A_P [{\rm \mu K^2}]$ & $27.0$ & $24.8$ & $<27.0$ & $<26.0$\\
\hline
$\chi^2_{\rm min}$ & $8128.4$ & $8124.0$ & $8126.2$ & $8125.6$\\
\hline
\hline
\end{tabular}
\caption{Similar constraints as in Table \ref{wvar}, but now also including the SNeIa data in the analysis.}
\label{snwvar}
\end{center}
\end{table*}

The standard $\Lambda$CDM cosmological model has some conceptual problems 
that are partially alleviated in some models in which the dark energy 
density varies slowly in time (and so weakly in space), \cite{Peebles:1988ab}.
Furthermore, observational constraints on cosmological parameters are 
model dependent, i.e., the observational estimate of a cosmological 
parameter, e.g., $\neff$, depends on the cosmological model used
to analyse the data. It is therefore of interest to examine the 
observational cosmological constraints on $\neff$ in a cosmological
model in which the dark energy density varies in time, such as that
of Ref.\ \cite{Peebles:1988ab}. This is a somewhat challenging task 
that we will leave for future work. However, to get an indication 
of what could be expected from such an analysis, we determine the 
observational constraints on $\neff$ in a cosmological model in which 
the time-evolving
dark energy density is parametrized by the XCDM parametrization (made
complete by assuming that the acoustic spatial inhomogeneities propagate 
at the speed of light) described above. Table \ref{wvar} shows the observational 
constraints derived under these assumptions. 

From Table \ref{wvar} we see that the MS prior changes the best fit $w$ in the 
standard case with $\neff=3.046$ to $w \sim -0.9$, with $w=-1$ off by one 
standard deviation. When both $w$ and $\neff$ are allowed to vary freely 
the geometrical degeneracy with $H_0$ makes the HST and MS $H_0$ priors 
much less effective. In this case the evidence for dark radiation is again 
significant: for the MS $H_0$ prior case we find $\neff =4.16\pm0.53$, 
and a dark energy equation of state parameter $w=-0.76\pm 0.10$, i.e., 
excluding a cosmological constant at more than two standard deviations. 
A scale-invariant HPYZ primordial spectrum with $n_s=1$ is fully consistent 
with both priors. While some of these values indicate significant tensions 
with the standard $\Lambda$CDM model, it is important to keep in mind the 
strong degeneracies between $\neff$, $H_0$ and $w$, as well as the fact 
that the XCDM parametrization used in the analysis has been arbitrarily 
completed to allow for an accounting of the evolution of density 
inhomogeneities.

In order to try to break these degeneracies, and derive more reliable 
constraints on the parameters, we perform a new analysis that also include 
the SDSS supernova Type Ia (SNeIa) apparent magnitude data, \cite{kessler}.
From Table \ref{snwvar} we see that the inclusion of the
SNeIa data bring the results back to the previous dichotomy:
the HST prior clearly shows a preference for $\neff>3.046$ while the MS 
prior results in a value of $\neff$ that is in much better agreement 
with the standard scenario. The constraints on the equation of state 
are $w=-0.967\pm0.075$ for the HST prior and $w=-0.946\pm0.076$ for  
the MS prior. The HPYZ spectrum with $n_s=1$ is again in tension with 
the observations for the MS $H_0$ prior at a little less than two 
standard deviations.

\section{Conclusions}

A ``standard'' cosmological model is starting to fall in 
place. Interestingly, recent data have provided some indication for an
unexpected new ``dark radiation'' component. In this brief paper 
we have again emphasized the important role played by the HST 
$H_0$ prior in establishing the statistical evidence for the 
existence of this dark radiation. We have also shown that with 
a new median statistics $H_0$ prior derived from 537 non-CMB 
$H_0$ measurements, there is no significant evidence for 
$\neff > 3.046$, consistent with the indications from other 
cosmological data. And it is probably not unreasonable to believe 
that the converse might also be true: with other cosmological 
data not inconsistent with $\neff = 3.046$, consistency of the 
smaller-scale CMB anisotropy data with the predictions of the 
$\Lambda$CDM model apparently demands $H_0 \sim 68 \mks$. 

We emphasize, however, that when the same data are analysed 
in the context of the somewhat arbitrarily-completed XCDM 
dark energy parametrization, they prefer $\neff>3.046$. It 
would be useful to see if this remains the case if, instead of
XCDM, a complete and consistent model, such as $\phi$CDM, is used 
in the analysis. 

We have shown that the HST $H_0$ prior is, at least partially,
responsible for the evidence supporting the existence of a new 
dark radiation component. However, future CMB anisotropy and galaxy 
clustering data, as well as a definitive determination of $H_0$,
will be needed to fully resolve this issue.

In particular, the future data expected from the Planck satellite
should be able to constrain independently the values of $H_0$ and
$\neff$, clarifying if the current tension between the HST
and CMB constraints on $H_0$ is due to a dark radiation component
or systematics in the data.

\section{Acknowledgments}
This work is supported by PRIN-INAF ``Astrono\-my probes fundamental physics", 
the Italian Space Agency through ASI contract Euclid-IC (I/031/10/0), 
DOE grant DEFG030-99EP41093, and NSF grant AST-1109275.

\end{document}